\documentclass[prd,showpacs,preprintnumbers,groupedaddress,superscriptaddress,11pt]{revtex4-1} 

\usepackage{amsthm,amsmath}
\newtheoremstyle{newplain}
{}
{}
{\it}
{}
{\bfseries}
{.}
{5pt}
{\thmname{#1}\hspace{5pt}\thmnumber{#2}\thmnote{\hspace{2pt}[{\normalfont #3}]}}
\theoremstyle{newplain}
\newtheorem{theorem}{Theorem}
\newtheorem{definition}{Definition}

\newtheorem{proposition}{Proposition}
\usepackage{cases}
\usepackage{ulem}
\usepackage{amsfonts}

\usepackage[dvipdfmx]{hyperref}
\hypersetup{%
 setpagesize=false,
 bookmarksnumbered=true,%
 bookmarksopen=true,%
 colorlinks=true,%
 linkcolor=blue,%
 citecolor=red}

\usepackage{color}
\usepackage{comment}

\begin{document}

\preprint{RUP-20-34}
\preprint{KEK-Cosmo-0268, KEK-TH-2280}

\title{Photon surfaces in less symmetric spacetimes}


\author{Yasutaka Koga}
\email[]{koga@rikkyo.ac.jp}
\affiliation{Department of Physics, Rikkyo University, Toshima, Tokyo 171-8501, Japan}
\author{Takahisa Igata}
\email[]{igata@post.kek.jp}
\affiliation{KEK Theory Center, 
Institute of Particle and Nuclear Studies, 
High Energy Accelerator Research Organization, Tsukuba 305-0801, Japan}
\author{Keisuke Nakashi}
\email[]{nakashi@rikkyo.ac.jp}
\affiliation{Department of Physics, Rikkyo University, Toshima, Tokyo 171-8501, Japan}


\date{\today}

\begin{abstract}
We investigate photon surfaces and their stability in a less symmetric spacetime, a general static warped product with a warping function acting on a Riemannian submanifold of codimension two.
We find a one-dimensional pseudopotential that gives photon surfaces as its extrema regardless of the spatial symmetry of the submanifold.
The maxima and minima correspond to unstable and stable photon surfaces, respectively.
It is analogous to the potential giving null circular orbits in a spherically symmetric spacetime.
We also see that photon surfaces indeed exist for the spacetimes which are solutions to the Einstein equation.
The parameter values for which the photon surfaces exist are specified.
As we show finally, the pseudopotential arises due to the separability of the null geodesic equation, and the separability comes from the existence of a Killing tensor in the spacetime.
The result leads to the conclusion that photon surfaces may exist even in a less symmetric spacetime if the spacetime admits a Killing tensor.
\end{abstract}


\maketitle


\tableofcontents

\section{Introduction}
\label{sec:introduction}
A photon surface was defined as the geometrical generalization of the photon sphere of the Schwarzschild spacetime by Claudel {\it et al.}~\cite{claudel}.
The surface is a hypersurface of a spacetime to which every initially tangent null geodesic remains tangent, and any global symmetry of the surface and the spacetime is not assumed in the definition.
Claudel {\it et al.}~\cite{claudel} proved that a photon surface is equivalent to a totally umbilic hypersurface, i.e., a hypersurface on which the trace-free part of the second fundamental form vanishes everywhere, in a spacetime of dimension four, and subsequently, Perlick~\cite{perlick} proved it in a spacetime of arbitrary dimensions.
The works enabled the geometrical analysis of a photon surface, and there have been various discoveries concerning a photon surface:
similarly to the black hole uniqueness theorems, uniqueness theorems of spacetimes possessing photon surfaces have been established in Refs.~\cite{cederbaum,cederbaum_maxwell,yazadjiev_psuniqueness,rogatko_psuniqueness};
in an accretion problem of radiation fluid, there is a correspondence, called {\it the sonic point/photon surface correspondence}, between the sonic points of the flow and photon surfaces~\cite{koga3,tsuchiya};
throats of pure-tensional thin shell wormholes and branes of the brane world model~\cite{randall,randall2} were found to be photon surfaces in Ref.~\cite{koga:psf-wht}.
\par
There are several examples of photon surfaces~\cite{claudel}.
In the Minkowski spacetime, timelike planes and single-sheeted hyperboloids are photon surfaces.
Photon surfaces, or we may call them photon spheres in this case, have been found for spherically symmetric black hole spacetimes that are solutions to the Einstein equation. 
Similarly, photon surfaces exist in the hyperbolically and planar symmetric counterparts of the spherically symmetric spacetime solutions~\cite{koga3}.
However, a photon surface does not exist for rotating vacuum black hole spacetimes such as the Kerr spacetime.
One may expect that, in spite of the definition without any explicit requirement of symmetry, photon surfaces exist only in highly symmetric spacetimes, specifically spacetimes of cohomogeneity one.
\par
However, Gibbons and Warnick found photon surfaces for the C-metric, which is the solution to the Einstein equation and of cohomogeneity two, and its generalizations including dilaton fields~\cite{gibbons_2016}.
This work revealed the existence of photon surfaces in a less symmetric spacetime of cohomogeneity more than one and allows us to expect that photon surfaces may exist regardless of the symmetry of the spacetime.
\par
In this paper, we investigate photon surfaces in a class of less or nonsymmetric spacetimes and discuss a structure that enables the spacetimes to have photon surfaces.
First, we show that in the class of spacetimes, the problem of finding a photon surface reduces to that of solving a one-dimensional equation given by a pseudopotential.
Second, we explicitly see that there exist photon surfaces in the electrovacuum cases of the spacetimes.
Finally, we show that the pseudopotential arises due to the separability of the null geodesic equation and conclude that the existence of a Killing tensor is crucial for the spacetimes to have photon surfaces in the present case.
\par
We consider a spacetime $(M,g)$ with the metric ansatz, 
\begin{equation}
\label{eq:metric-arbitrary}
g=-f(r)dt^2+h(r)dr^2+r^2\gamma_{ij}\left(x\right)dx^idx^j,
\end{equation}
where we assume $f(r),\ h(r)>0$, and investigate photon surfaces of constant $r$.~\footnote{
In the current paper, we assume $f(r),\ h(r)>0$ for simplicity.
The condition $h(r)>0$ is necessary for a hypersurface of $r=\mathrm{const.}$ to be timelike.
The condition $f(r)>0$ is necessary for the metric $g$ to have Lorentzian signature if we require the metric $\gamma_{ij}$ to have Riemannian signature.
One can instead assume that $f(r)<0$, $h(r)>0$, and $\gamma_{ij}$ is Lorentzian.
Since our analysis does not depend on the signature, the analysis in that case should be parallel.}
The spacetime dimension is $D\ge3$, and the $(D-2)$-dimensional Riemannian metric $\gamma_{ij}(x)=\gamma_{ij}(x^1,x^2,...,x^{D-2})$ are arbitrary.
The spacetime can be derived as the generic form of a some class of warped product.
See Appendix~\ref{app:warped-product} for the derivation of Eq.~(\ref{eq:metric-arbitrary}).
The spacetime is static but has no spatial symmetry in general.
If $\gamma_{ij}(x)$ is the metric of the unit $(D-2)$-sphere, $g$ is the metric of a general static spherically symmetric spacetime, and therefore, the photon surfaces of constant $r$ are what we usually call photon spheres.
\par
This paper is organized as follows.
In Sec.~\ref{sec:photon-surface}, we review a photon surface and its stability.
In Sec.~\ref{sec:const-r-psf}, we define the {\it $r$-photon surface} and derive a one-dimensional pseudopotential $V(r)$, which allows us to find the photon surface and to analyze its stability easily.
In Sec.~\ref{sec:vacuum-solution}, we specifically consider the $\Lambda$-electrovacuum solutions to the Einstein equation with the ansatz~(\ref{eq:metric-arbitrary}) and specify the parameter ranges in which the $r$-photon surfaces exist.
In Sec.~\ref{sec:killing-tensor}, we see that a Killing tensor is responsible for introducing the one-dimensional pseudopotential $V(r)$,
and therefore, the less symmetric spacetime admits $r$-photon surfaces in several cases.
Section~\ref{sec:conclusion} is devoted to the conclusion.
We use units in which $G=1$ and $c=1$. 
The roman indices $a,b,...$ of tensors are the abstract indices~\cite{textbook:wald}.

\section{Photon surface and stability}
\label{sec:photon-surface}
Here we review a photon surface and its stability.
\subsection{Photon surface}
A photon surface is a hypersurface on which every null geodesic initially tangent to it remains tangent.
It is defined by Claudel {\it et al.}~\cite{claudel}:
\begin{definition}[Photon surface]
\label{definition:photonsurface}
A photon surface of a spacetime $(M, {g})$ is an immersed, nowhere-spacelike
hypersurface $S$ of $(M, {g})$ such that, for every point $p\in S$ and every null vector ${k}\in T_pS$, there exists a null geodesic $\gamma\colon (-\epsilon,\epsilon) \to M$ of $(M, {g})$ such that $\dot{\gamma}(0) ={k}, |\gamma|\subset S$.
\end{definition}
There exists an equivalent condition for a timelike hypersurface to be a photon surface. 
This is summarized as the following theorem proven in four dimensions by Claudel {\it et al.}~\cite{claudel} and in arbitrary dimensions by Perlick~\cite{perlick}:
\begin{theorem}[Claudel-Virbhadra-Ellis (2001), Perlick (2005)]
\label{theorem:equivalent-condition}
Let $S$ be a timelike hypersurface of a spacetime $(M,g)$.
Let $n$ be a unit normal to $S$ and let $h_{ab}=g_{ab}-n_an_b$ be the induced metric on $S$.
Let $\chi_{ab}=h^c_a\nabla_cn_b$ be the second fundamental form of $S$ and let $\sigma_{ab}=\chi_{ab}-[h^{cd}\chi_{cd}/(D-1)]h_{ab}$ be the trace-free part of $\chi_{ab}$, where $D$ is the dimension of $(M,g)$.
Then $S$ is a photon surface if and only if it is totally umbilic, i.e.,
\begin{equation}
\label{eq:umbilic}
\sigma_{ab}=0\;\forall p\in S.
\end{equation}
\end{theorem}
\subsection{Stability}
The stability of null geodesics on a photon surface is defined in Ref.~\cite{koga:psfstability}.
It represents whether or not a null geodesic $\gamma$ on $S$ is attracted toward $S$ if perturbed in the direction normal to $S$:
\begin{definition}[Stability of null geodesics on a photon surface]
\label{definition:stability-gamma}
Let $S$ be a timelike photon surface of $(M,g)$ and $n$ be a unit normal vector to $S$.
Let $\gamma$ be a null geodesic on $S$ passing through a point $p\in S$ and $k$ be the tangent vector to $\gamma$.
Let $X$ be the deviation vector of $\gamma$ satisfying the condition,
\begin{equation}
\left.X\right|_p\propto \left.n\right|_p.
\end{equation}
The null geodesic $\gamma$ is said to be stable, unstable, and marginally stable at $p$ if the acceleration scalar $a:=g\left(X,\nabla_k\left(\nabla_kX\right)\right)$ satisfies
\begin{equation}
\label{eq:stability-orbit}
\left.a\right|_p<0,\ \ >0,\ \ \ and\ =0,
\end{equation}
respectively.
\end{definition}
Two generic formulas for the stability are given in Ref.~\cite{koga:psfstability}.
One is written in terms of the Riemann curvature and can be easily derived from Eq.~\eqref{eq:stability-orbit}.
The other one is given in terms of the trace-free part of the second fundamental form and is convenient for the current purpose:
\begin{proposition}
\label{proposition:stability-shear}
Let $S$ be a timelike photon surface and $\left\{S_y\right\}\ni S_0:=S$ be the Gaussian normal foliation~\cite{koga:psfstability} with respect to $S$.
Let  $n$ and $\sigma_{ab}$ be the unit normal field and the trace-free part of the second fundamental form of each $S_y$, respectively.
A null geodesic $\gamma$ on $S$ with the tangent $k$ is said to be stable, unstable, and marginally stable at $p\in S$ if and only if
\begin{equation}
\label{eq:stabilitycondition-shear}
k^ak^b
\left.\nabla_n\sigma_{ab}
\right|_p<0,\;>0,\; and\; =0,
\end{equation}
respectively.
\end{proposition}
{\it The Gaussian normal foliation} $\left\{S_y\right\}$ in the above is a foliation satisfying the condition,
\begin{equation}
\label{eq:gaussian-foliation}
dn=0,
\end{equation}
for the unit normal field $n$ to each $S_y$.
This is the foliation obtained by taking the Gaussian normal coordinates since $n$ is the geodesic tangent orthogonal to the surfaces~\cite{koga:psfstability}.
Note that the notion of the stability does not change under the flip of the normal field, $n\to -n$.
The second fundamental form $\chi_{ab}$ of $S_y$ is given by $\chi_{ab}=h^c_a\nabla_cn_b$ with the induced metric $h_{ab}=g_{ab}-n_an_b$, and therefore, its normal derivative $\nabla_n\chi_{ab}=n^d\nabla_d\left(h^c_a\nabla_cn_b\right)$ is invariant under the flip.
Similarly, $\nabla_n\sigma_{ab}$ is also invariant.
\par
Note that even if a null geodesic $\gamma$ on a photon surface $S$ is stable at a point $p\in|\gamma|$, it can be unstable at another point $q\in|\gamma|$ in general.
If every null geodesic on the photon surface $S$ is stable (unstable) at every point, the photon surface $S$ itself is said to be stable (unstable) as follows:
\begin{definition}[Stability of a photon surface]
\label{definition:stability-psf}
A timelike photon surface $S$ is said to be
\begin{itemize}
\item stable if every null geodesic $\gamma$ on $S$ is stable or marginally stable at every point $p\in|\gamma|$,
\item strictly stable if every null geodesic $\gamma$ on $S$ is stable at every point $p\in|\gamma|$,
\item unstable if every null geodesic $\gamma$ on $S$ is unstable or marginally stable at every point $p\in|\gamma|$,
\item strictly unstable if every null geodesic $\gamma$ on $S$ is unstable at every point $p\in|\gamma|$, and
\item marginally stable if every null geodesic $\gamma$ on $S$ is marginally stable at every point $p\in|\gamma|$.
\end{itemize}
\end{definition}

\section{$r$-photon surface}
\label{sec:const-r-psf}
Here, we consider a spacetime $(M,g)$ with the metric~\eqref{eq:metric-arbitrary}.
We investigate {\it an $r$-photon surface} defined below and derive its stability condition in what follows.
\subsection{Definition}
An $r$-photon surface is a photon surface of a hypersurface $r=\mathrm{const.}$ in a spacetime with the metric given by Eq.~\eqref{eq:metric-arbitrary}:
\begin{definition}[$r$-photon surface]
\label{definition:r-photonsurface}
Let $(M,g)$ be a spacetime with the metric~(\ref{eq:metric-arbitrary}).
A hypersurface $S_r$ defined by
\begin{equation}
S_r:=\left\{p\in M|r=\mathrm{const.} \right\}
\end{equation}
is called an $r$-photon surface if it is a photon surface.
\end{definition}
\subsection{Condition for an $r$-photon surface}
\label{sec:psf-condition}
For general $S_r$, its unit normal is given by
\begin{equation}
\label{eq:unit-normal-sr}
n=\sqrt{h}dr.
\end{equation}
The induced metric of $S_r$ is given by
\begin{equation}
h_{ab}=-f(dt)_a(dt)_b+r^2\gamma_{ij}(dx^i)_a(dx^j)_b.
\end{equation}
The second fundamental form of $S_r$ and its trace-free part are given by
\begin{equation}
\chi_{ab}=\frac{1}{2}h^{-1/2}\left[-f'(dt)_a(dt)_b+2r\gamma_{ij}(dx^i)_a(dx^j)_b\right]
\end{equation}
and
\begin{equation}
\label{eq:sigma}
\sigma_{ab}=-\frac{1}{2(D-1)}\frac{(fr^{-2})'}{fr^{-2}}h^{-1/2}\left[(D-1)f_{ab}+h_{ab}\right],
\end{equation}
respectively, where $f_{ab}:=f(r)(dt)_a(dt)_b$.
Theorem~\ref{theorem:equivalent-condition} gives the condition for the $r$-photon surface:
\begin{proposition}
\label{proposition:r-psf}
\label{proposition:const-r-psf-condition}
A timelike hypersurface $S_r$ is an $r$-photon surface if and only if
\begin{equation}
\label{eq:r-psf}
(fr^{-2})'=0
\end{equation}
at $r$.
\end{proposition}
\subsection{Stability condition}
Let $\left\{S_r\right\}$ be a foliation of the spacetime $(M,g)$ with Eq.~(\ref{eq:metric-arbitrary}).
Since the unit normal $n$ given in Eq.~\eqref{eq:unit-normal-sr} satisfies the condition~\eqref{eq:gaussian-foliation}, it is a Gaussian normal foliation.
With $\sigma_{ab}$ defined on each $S_r$, we calculate $\nabla_n\sigma_{ab}$.
For a radius $r=r_{\mathrm{p}}$ such that $S_{r_{\mathrm{p}}}$ is an $r$-photon surface, we have
\begin{equation}
\left.\nabla_n\sigma_{ab}\right|_{r=r_{\mathrm{p}}}=\left.-\frac{1}{2(D-1)}\frac{(fr^{-2})''}{fr^{-2}}h^{-1}\left[(D-1)f_{ab}+h_{ab}\right]\right|_{r=r_{\mathrm{p}}}
\end{equation}
by using Eq.~(\ref{eq:r-psf}).
For any null vector $k\in T_pS_{r_{\mathrm{p}}}$,
\begin{equation}
k^ak^b\left.\nabla_n\sigma_{ab}\right|_{r=r_{\mathrm{p}}}=-\left.\frac{1}{2}(fr^{-2})''r^2h^{-1}(k^t)^2\right|_{r=r_{\mathrm{p}}}.
\end{equation}
From Proposition~\ref{proposition:stability-shear}, this equation implies that any null geodesic on an $r$-photon surface is stable, unstable, and marginally stable at any point if $(fr^{-2})''>0$, $<0$, and $=0$, respectively.
Then, according to Definition~\ref{definition:stability-psf}, the stability condition of an $r$-photon surface is obtained as follows:
\begin{proposition}
\label{proposition:stability}
A timelike $r$-photon surface $S_{r_{\mathrm{p}}}$ is strictly stable, strictly unstable, and marginally stable if and only if
\begin{equation}
\label{eq:stability}
\left.(fr^{-2})''\right|_{r=r_{\mathrm{p}}}>0,\ <0,\ and\ =0,
\end{equation}
respectively.
\end{proposition}
The $r$-photon surfaces are classified into only the three types, strictly stable, strictly unstable, and marginally stable ones, which are not overlapped each other.
Therefore, we simply call them stable, unstable, marginally stable $r$-photon surfaces in the following.
\subsection{Pseudopotential}
Here we define {\it the pseudopotential}
\begin{equation}
\label{eq:pp}
V(r):=f(r)r^{-2},
\end{equation}
which is useful for investigating $r$-photon surfaces.
From Proposition~\ref{proposition:r-psf} and~\ref{proposition:stability}, the conditions for an $r$-photon surface are given as follows:
\begin{proposition}
\label{proposition:pseudo-potential}
Suppose that $h(r_{\mathrm{p}})>0$ at $r=r_{\mathrm{p}}$, and therefore, $S_{r_\mathrm{p}}$ is timelike.
Then $S_{r_{\mathrm{p}}}$ is an $r$-photon surface if and only if
\begin{equation}
\label{eq:V'=0}
V'(r_{\mathrm{p}})=0.
\end{equation}
It is stable, unstable, and marginally stable if and only if
\begin{equation}
\label{eq:stability-V}
V''(r_{\mathrm{p}})>0,\ <0,\ and\ =0,
\end{equation}
respectively.
\end{proposition}
$r$-photon surfaces appear as the extrema of $V(r)$.
Unstable and stable $r$-photon surfaces correspond to the local maxima and minima of $V(r)$, respectively.

\section{$r$-photon surfaces in vacuum spacetimes}
\label{sec:vacuum-solution}
We see that $r$-photon surfaces exist in spacetimes of solutions to the Einstein equation.
\subsection{Electrovacuum spacetime with the cosmological constant}
The ansatz~(\ref{eq:metric-arbitrary}) gives the solution to the electrovacuum Einstein equation with the cosmological constant given by the action,
\begin{equation}
S=\int d^Dx  \sqrt{-g} (R-2\Lambda-F^{ab}F_{ab}),
\end{equation}
where $R$, $\Lambda$, and $F_{ab}$ are the Ricci scalar of $(M,g)$, the cosmological constant, and the field strength of the electromagnetic field, respectively.
From the field equation with the ansatz of the electromagnetic field,
\begin{equation}
F=\sqrt{\frac{(D-2)(D-3)}{2}} \frac{Q}{r^{D-2}} dt \wedge dr,
\end{equation}
the metric components of the ansatz~(\ref{eq:metric-arbitrary}) are given as follows~\cite{kodama_2004}:
\begin{equation}
\label{eq:vacuum-solution}
f(r)=h^{-1}(r)=k-\frac{2\Lambda}{(D-2)(D-1)} r^2-\frac{2M}{r^{D-3}}+\frac{Q^2}{r^{2(D-3)}},
\end{equation}
where $k$ is a constant relevant to $\gamma_{ij}$, $M$ is the mass parameter, and $Q$ is the electric charge parameter.
The $(D-2)$-dimensional Riemannian submanifold $\Sigma$ with the metric $\gamma_{ij}$ is an arbitrary Einstein manifold with the relation to $k$, $\mathcal{R}_{ij}=(D-3)k \gamma_{ij}$, where $\mathcal{R}_{ij}$ is the Ricci curvature associated with $\gamma_{ij}$. 
The metric is invariant under an appropriate simultaneous scaling of the coordinates $t, r$ and the parameters $k,M,Q,\Lambda$.
For nonzero $k$, we can scale it so that $k=\pm1$.
\par
The submanifold $(\Sigma,\gamma)$ of dimension $N=D-2$ can be a constant curvature space because a constant curvature space is an Einstein manifold for any $N$.
For $N<4$, $(\Sigma,\gamma)$ is always a constant curvature space.
For $N=1$, although the Ricci curvature is not defined, the metric can always be written in the form of a flat space, $\gamma_{ij}(x)dx^idx^j=dl^2$.
For $N=2,3$, the condition $\mathcal{R}_{ij}=(D-3)k \gamma_{ij}$ implies that $(\Sigma,\gamma)$ is a constant curvature space because any sectional curvature is then constant $k$.
For $N\ge4$, various nontrivial Einstein manifolds have been found.
The variety of Einstein manifolds in higher $N$ is due to the degrees of freedom of the Weyl curvature.
See Refs.~\cite{jensen,fine,alekseevsky,grajales,textbook:einsteinmanifold} for examples of nontrivial Einstein manifolds.
\par
The solution~(\ref{eq:vacuum-solution}) would be a less symmetric spacetime in the sense that the $(D-2)$-dimensional submanifold $(\Sigma,\gamma)$ can be less symmetric than the maximal if $D\ge6$.
If there exist Einstein manifolds without any spatial Killing vectors, the spacetime can be a static electrovacuum spacetime with only the timelike Killing vector $\partial_t$.
\subsection{Pseudopotential}
Let us focus on a timelike hypersurface $S_{r_{\mathrm{p}}}$. 
The timelike condition of $S_{r_{\mathrm{p}}}$ is given by 
\begin{equation}
h(r_\mathrm{p})>0.
\end{equation}
Or, according to Eq.~\eqref{eq:vacuum-solution}, it is equivalent to
\begin{align}
\label{eq:tcond}
f(r_\mathrm{p})>0.
\end{align}
For $S_{r_{\mathrm{p}}}$ to be an $r_{\mathrm{p}}$-photon surface, 
the radius $r_{\mathrm{p}}$ must be a real positive solution to Eq.~\eqref{eq:V'=0}.
Using the explicit form of $V(r)$, 
\begin{align}
V(r)=-\frac{2\Lambda}{(D-2)(D-1)} +\frac{k}{r^2}-\frac{2M}{r^{D-1}}+\frac{Q^2}{r^{2(D-2)}},
\end{align}
Eq.~\eqref{eq:V'=0} reduces to
\begin{align}
\label{eq:rpeq}
k r_{\mathrm{p}}^{2(D-3)}-(D-1) M r_{\mathrm{p}}^{D-3}+(D-2) Q^2=0.
\end{align}
Once we find an $r_{\mathrm{p}}$-photon surface, we can 
determine the stability of $S_{r_{\mathrm{p}}}$ by the sign of 
\begin{align}
\label{eq:V''}
V''(r_{\mathrm{p}})=\frac{4\:\!(D-3)}{r_{\mathrm{p}}^{4}} \left[\:\!
\frac{(D-1)M}{2\:\!r_{\mathrm{p}}^{D-3}}-k
\:\!\right].
\end{align}

Let us focus the case $D=3$. Then Eqs.~\eqref{eq:tcond}, \eqref{eq:rpeq}, and \eqref{eq:V''}
 reduce to $f(r_{\mathrm{p}})=-\Lambda r_{\mathrm{p}}^2>0$, 
$k-2M+Q^2=0$, and $V''(r_{\mathrm{p}})=0$, respectively. 
Therefore, we conclude that only for $Q^2=2M-k\geq 0$ and $\Lambda<0$ 
(i.e., 3D anti--de Sitter spacetime), 
a hypersurface $S_{r_\mathrm{p}}$ is a timelike $r_{\mathrm{p}}$-photon surface 
at any $r_{\mathrm{p}}$ and is marginally stable.

We focus on the case $D\geq 4$ in the what follows. 
In each case, $k=0$ in Sec.~\ref{sec:k=0}, $k\neq0 \ \&\ M=0$ in Sec.~\ref{sec:kn0M0}, 
and $k\neq0\ \&\ M\neq0$ in Sec.~\ref{eq:kn0Mn0}, 
we apply the following procedure to show the existence of 
a timelike $r_{\mathrm{p}}$-photon surface: 
First, we find a solution to Eq.~\eqref{eq:rpeq} 
and restrict parameters to the range where the solution 
is real and positive.
The timelike condition~\eqref{eq:tcond} further restricts the allowed range of a 
dimensionless cosmological constant, 
\begin{align}
\lambda:=\Lambda \:\!|M|^{2/(D-3)}.
\end{align}
Evaluating the sign of Eq.~\eqref{eq:V''}, 
we determine the stability of $S_{r_{\mathrm{p}}}$ 
by Proposition~\ref{proposition:pseudo-potential}.
Here for later convenience, we introduce a dimensionless charge for $M\neq 0$, 
\begin{align}
 &q:= \frac{|Q|}{|M|}.
\end{align}
The results are summarized in Table~\ref{table:k-zero} for $k=0$, Table~\ref{table:k-plus} for $k=+1$, and Table~\ref{table:k-minus} for $k=-1$.
Similar analysis is found in the context of the stability of a thin shell wormhole throat~\cite{kokubu}.

\subsection{$k=0$}
\label{sec:k=0}
Suppose that $k=0$. Then Eqs.~\eqref{eq:rpeq} and \eqref{eq:V''} reduce to
\begin{align}
\label{eq:PSk0}
&(D-1)M r_{\mathrm{p}}^{D-3}-(D-2) Q^2=0,
\\
&V''(r_{\mathrm{p}})=(D-3)(D-1)\frac{2M}{r_{\mathrm{p}}^{D+1}}.
\end{align}
Note that if $S_{r_{\mathrm{p}}}$ is a timelike $r_{\mathrm{p}}$-photon surface, 
then it is stable for $M>0$, unstable for $M<0$, and marginally stable for $M=0$ 
because the sign of $M$ coincides with that of $V''(r_{\mathrm{p}})$. 
In the followings, we consider each case, $M=0$ and $M\neq 0$, separately.

\subsubsection{$k=0 \  \&\ M=0$}
Suppose that $k=0$ and $M=0$. 
Then Eq.~\eqref{eq:PSk0} leads to $Q=0$. 
Hence, $V$ becomes constant, 
$V=-2\Lambda/[(D-2)(D-1)]$, and satisfies
$V'(r_{\mathrm{p}})=0$ and $V''(r_{\mathrm{p}})=0$ for any value of $r_{\mathrm{p}}$. 
The timelike condition of $S_{r_\mathrm{p}}$ in Eq.~\eqref{eq:tcond} reduces to 
$f(r_{\mathrm{p}})=-2\Lambda r_{\mathrm{p}}^2/[(D-2)(D-1)]>0$, and therefore,
\begin{align}
\label{eq:lambdaads}
\Lambda<0.
\end{align}
Finally we conclude that only for $k=0$, $M=0$, $Q=0$, and $\Lambda<0$, 
a timelike $r_{\mathrm{p}}$-photon surface exists at any $r_{\mathrm{p}}>0$ 
 and is marginally stable.

\subsubsection{$k=0 \  \&\ M\neq 0$}
Suppose that $k=0$ and $M\neq 0$. Then we obtain the solution to Eq.~\eqref{eq:PSk0} in the form
\begin{align}
\label{eq:rpk0Mn0}
r_{\mathrm{p}}^{D-3}=\frac{D-2}{D-1} \frac{Q^2}{M}. 
\end{align}
The positivity of $r_{\mathrm{p}}$ requires $Q\neq 0$ and $M>0$, 
and thus, $V''(r_{\mathrm{p}})>0$. 
The timelike condition of $S_{r_\mathrm{p}}$ in Eq.~\eqref{eq:tcond} 
leads to a negative upper bound for $\lambda$, 
\begin{align}
\label{eq:lambdaJ}
\lambda<
-\frac{(D-3)(D-1)^2}{2\:\!(D-2)} \left(
\frac{D-1}{D-2}
\right)^{2/(D-3)} q^{-2(D-1)/(D-3)}<0. 
\end{align}
Finally we conclude that only for $k=0$, $M>0$, $Q\neq 0$, and Eq.~\eqref{eq:lambdaJ}, 
a timelike $r_{\mathrm{p}}$-photon surface exists at the radius~\eqref{eq:rpk0Mn0} and 
is stable. 

\subsection{$k\neq 0 \ \&\ M=0$}
\label{sec:kn0M0}
Suppose that $k\neq0$ and $M=0$. Then Eq.~\eqref{eq:rpeq} reduces to 
\begin{align}
kr_{\mathrm{p}}^{2(D-3)}+(D-2)Q^2=0.
\end{align}
The positivity of $r_{\mathrm{p}}$ requires $Q\neq 0$ and $k=-1$. 
The positive branch takes the form
\begin{align}
\label{eq:rpk-1M0}
r_{\mathrm{p}+}^{D-3}=\sqrt{D-2} \:\!|\:\!Q\:\!|
\end{align}
and leads to $V''(r_{\mathrm{p}+})=4(D-3)/r_{\mathrm{p}+}^4>0$ from Eq.~\eqref{eq:V''}.
The timelike condition~\eqref{eq:tcond} gives a negative upper bound for $\Lambda$, 
\begin{align}
\label{eq:lambdakn0Mn0}
\Lambda <
-  \frac{(D-3)(D-1)}{2\:\!(D-2)^{1/(D-3)}} |Q|^{-2/(D-3)}<0.
\end{align}
Finally we conclude that 
only for $k=-1$, $M=0$, $Q\neq0$, and Eq.~\eqref{eq:lambdakn0Mn0}, 
a timelike $r_{\mathrm{p}}$-photon surface exists at the radius~\eqref{eq:rpk-1M0} 
and is stable.

\subsection{$k\neq 0 \  \&\ M\neq 0$}
\label{eq:kn0Mn0}
Suppose that $k\neq 0$ and $M\neq 0$. Then Eq.~\eqref{eq:rpeq} has roots 
\begin{align}
\label{eq:roots}
r_{\mathrm{p}\pm}^{D-3}=\frac{D-1}{2\:\!k} M(1\pm \gamma),
\end{align}
where 
\begin{align}
&\gamma:= \sqrt{1-k \frac{q^2}{q_{\mathrm{c}}^2}},
\\
&q_{\mathrm{c}}:= \frac{D-1}{2\sqrt{D-2}}.
\end{align}
Using these roots, $V''(r_{\mathrm{p}})$ in Eq.~\eqref{eq:V''} is formally written as 
\begin{align}
\label{eq:V"kn0Mn0}
V''(r_{\mathrm{p}\pm})=\mp (D-3)(D-1) \frac{2M\gamma}{r_{\mathrm{p}\pm}^{D+1}},
\end{align}
where $V''(r_{\mathrm{p}+})$ corresponds to the upper sign in the right-hand side and vice versa. 
In the followings, we consider 
each case, $k=1$ and $k=-1$, separately. 

\subsubsection{$k=1$}
Suppose that $k=1$. Then $\gamma$ is restricted 
to the range $0\leq \gamma=(1-q^2/q_{\mathrm{c}}^2)^{1/2} \leq 1$ 
(i.e., $0\leq q\leq q_{\mathrm{c}}$). 
First, let us focus on the case $\gamma=1$ (i.e., uncharged case, $q=0$). 
The branch $r_{\mathrm{p}-}$ of the roots~\eqref{eq:roots} vanishes, and therefore, 
here is no photon surface. 
On the other hand, if $M>0$, then the branch
\begin{align}
\label{eq:rpk1}
r_{\mathrm{p}+}^{D-3}=(D-1) M
\end{align}
is positive definite, and $V''(r_{\mathrm{p}+})<0$ holds. 
The timelike condition~\eqref{eq:tcond} provides 
a positive upper bound of $\lambda$, 
\begin{align}
\label{eq:lambdak1}
\lambda<
\frac{(D-3)(D-2)}{2\:\!(D-1)^{2/(D-3)}}.
\end{align}
Finally we conclude that only for $k=1$, $M>0$, $q=0$, and Eq.~\eqref{eq:lambdak1}, 
a timelike $r_{\mathrm{p}+}$-photon surface exists at the radius~\eqref{eq:rpk1} 
and is unstable. 

Next, we focus on the case $\gamma=0$ (i.e., $q=q_{\mathrm{c}}$). 
The roots~\eqref{eq:roots} are degenerate as  
\begin{align}
\label{eq:rpk1qc}
r_{\mathrm{p}}^{D-3}=\frac{D-1}{2}M.
\end{align}
The positivity of $r_{\mathrm{p}}$ requires $M>0$. 
Since $\gamma=0$, we have 
$V''(r_{\mathrm{p}})=0$ from Eq.~\eqref{eq:V"kn0Mn0}. 
The timelike condition~\eqref{eq:tcond} provides 
a positive upper bound of $\lambda$,
\begin{align}
\label{eq:lambdak1qc}
\lambda < \frac{(D-3)^2}{2} \left(\frac{2}{D-1}\right)^{2/(D-3)}.
\end{align}
Finally we conclude that only for $k=1$, $M>0$, $q=q_{\mathrm{c}}$, and Eq.~\eqref{eq:lambdak1qc}, 
a timelike $r_{\mathrm{p}}$-photon surface exists at the radius~\eqref{eq:rpk1qc} and 
is marginally stable. 

Next, we focus on the case $0<\gamma<1$ (i.e., $0<q<q_{\mathrm{c}}$). 
If $M<0$, both roots~\eqref{eq:roots} are negative and hence unsuitable. 
Suppose that $M>0$. Then the roots satisfy
$r_{\mathrm{p}+}>r_{\mathrm{p}-}>0$. 
For each branch, $V''(r_{\mathrm{p}+})<0$ and $V''(r_{\mathrm{p}-})>0$. 
The timelike condition~\eqref{eq:tcond} reduces to 
upper bounds of $\lambda$, 
\begin{align}
\label{eq:lambdak1q01}
\lambda< \lambda_\pm(D, q):=
\frac{D-3}{2}
\left[\:\!
\frac{2}{(D-1)(1\pm \gamma)}
\:\!\right]^{2/(D-3)}
\left(
D-1
-\frac{2}{1\pm \gamma}
\right).
\end{align}
Note that $\lambda_+>0$, and on the other hand, $\lambda_-\geq 0$ for $1\leq q<q_{\mathrm{c}}$ and 
$\lambda_-<0$ for $0<q<1$. 
Finally we conclude that only for $k=1$, $M>0$, $0<q<q_{\mathrm{c}}$, and Eq.~\eqref{eq:lambdak1q01}, 
a timelike $r_{\mathrm{p}+}$-photon surface exists at the radius $r_{\mathrm{p}+}=(D-1)(1+\gamma)M/2$ and is unstable, and a timelike $r_{\mathrm{p}-}$-photon surface exists at the radius $r_{\mathrm{p}-}=(D-1)(1-\gamma)M/2$ and is stable.

\subsubsection{$k=-1$}
For $k=-1$, $\gamma$ is restricted to the range $\gamma=(1+q^2/q_{\mathrm{c}}^2)^{1/2}\geq 1$. 
We consider 
each case, $M>0$ and $M<0$, separately. 
 
Suppose that $k=-1$ and $M>0$. 
For $\gamma=1$ (i.e., uncharged case, $q=0$), 
the roots~\eqref{eq:roots} satisfy 
$r_{\mathrm{p}-}^{D-3} =0> r_{\mathrm{p}+}^{D-3}$, and hence, 
here are no photon surfaces. 
Now, we focus on the case $\gamma>1$ (i.e., $q\neq 0$). 
The branch $r_{\mathrm{p}-}$ of the roots~\eqref{eq:roots} becomes  
\begin{align}
\label{eq:rpk-1qn0}
r_{\mathrm{p}-}^{D-3}=\frac{D-1}{2}M (\gamma-1)>0,
\end{align}
while the other branch is unsuitable because $r_{\mathrm{p}+}^{D-3}<0$. 
We find that $V''(r_{\mathrm{p}-})>0$ from Eq.~\eqref{eq:V"kn0Mn0}. 
The timelike condition~\eqref{eq:tcond} provides a negative upper bound of $\lambda$,
\begin{align}
\label{eq:lambdak-1qn0}
\lambda< 
-\frac{D-3}{2} \left[\:\!
\frac{2}{(D-1)(\gamma-1)}
\:\!\right]^{2/(D-3)}\left(
D-1+\frac{2}{\gamma-1} 
\right)<0. 
\end{align}
Finally we conclude that only for $k=-1$, $M>0$, $q\neq 0$, and Eq.~\eqref{eq:lambdak-1qn0}, 
a timelike $r_{\mathrm{p}-}$-photon surface exists at the radius~\eqref{eq:rpk-1qn0} and is 
stable. 

Suppose that $k=-1$ and $M<0$. 
The branch $r_{\mathrm{p}+}$ of the roots~\eqref{eq:roots} becomes 
\begin{align}
\label{eq:rpk-1M-}
r_{\mathrm{p}+}^{D-3}=\frac{D-1}{2} |M| (1+\gamma),
\end{align}
while the other branch is unsuitable because $r_{\mathrm{p}-}^{D-3}<0$. 
We find that $V''(r_{\mathrm{p}+})>0$ from Eq.~\eqref{eq:V"kn0Mn0}. 
The timelike condition~\eqref{eq:tcond} provides a
negative upper bound of $\lambda$, 
\begin{align}
\label{eq:lambdak-1qn0M-}
\lambda<
-\frac{D-3}{2} \left[\:\!
\frac{2}{(D-1)(1+\gamma)}
\:\!\right]^{2/(D-3)} \left(
D-1-\frac{2}{1+\gamma}
\right)<0. 
\end{align}
Finally we conclude that only for $k=-1$, $M<0$, and Eq.~\eqref{eq:lambdak-1qn0M-}, 
a timelike $r_{\mathrm{p}+}$-photon surface exists at the radius~\eqref{eq:rpk-1M-}
and is stable.

\begin{table}[t]
\begin{tabular}{cccc}
\hline
\hline
&$M>0$&$M=0$&$M<0$
\\
\hline
$Q=0$&$\not{\exists}$&$\forall r$, Eq.~\eqref{eq:lambdaads}, marginally stable, &$\not{\exists}$
\\
$Q\neq0$&Eqs.~\eqref{eq:rpk0Mn0} and \eqref{eq:lambdaJ}, stable&$\not{\exists}$&$\not{\exists}$
\\
\hline\hline
\end{tabular}
\caption{$k=0$}
\label{table:k-zero}
\end{table}

\begin{table}[t]
\begin{tabular}{cccc}
\hline
\hline
&$M>0$&$M=0$&$M<0$
\\
\hline
$q=0$&Eqs.~\eqref{eq:rpk1} and \eqref{eq:lambdak1}, unstable&$\not{\exists}$&$\not{\exists}$
\\
$0<q<q_{\mathrm{c}}$
&Eqs.~\eqref{eq:roots} and \eqref{eq:lambdak1q01} (upper branch), unstable
&$\not{\exists}$&$\not{\exists}$\\ 
&Eqs.~\eqref{eq:roots} and \eqref{eq:lambdak1q01} (lower branch), stable~~~~~&&
\\
$q=q_{\mathrm{c}}$&Eqs.~\eqref{eq:rpk1qc} and \eqref{eq:lambdak1qc}, marginally stable&$\not{\exists}$&$\not{\exists}$\\
\hline\hline
\end{tabular}
\caption{$k=1$}
\label{table:k-plus}
\end{table}

\begin{table}[t]
\begin{tabular}{cccc}
\hline
\hline
&$M>0$&$M=0$&$M<0$
\\
\hline
$Q=0$&$\not{\exists}$& $\not{\exists}$&Eqs.~\eqref{eq:rpk-1M-} and \eqref{eq:lambdak-1qn0M-}, stable
\\
$Q\neq 0$&Eqs.~\eqref{eq:rpk-1qn0} and \eqref{eq:lambdak-1qn0}, stable
&Eqs.~\eqref{eq:rpk-1M0} and \eqref{eq:lambdakn0Mn0}, stable
&Eqs.~\eqref{eq:rpk-1M-} and \eqref{eq:lambdak-1qn0M-}, stable\\
\hline\hline
\end{tabular}
\caption{$k=-1$}
\label{table:k-minus}
\end{table}

\section{Photon surface and the Killing tensor}
\label{sec:killing-tensor}
We have seen that, regardless of the spatial symmetry of the spacetime under consideration, the $r$-photon surfaces are given by the equation of the single variable $r$, Eq.~\eqref{eq:V'=0}, and further exist in several electrovacuum cases.
However, the reason for this result is quite nontrivial from the geometrical analysis based on Theorem~\ref{theorem:equivalent-condition}, which we have adopted so far.
In this section, we return to the null geodesic equation and see that its separability in $r$ is closely related to the appearance of the pseudopotential $V(r)$ and the existence of $r$-photon surfaces.
We finally conclude that the less or nonsymmetric spacetime can have the $r$-photon surface in several cases due to the existence of a rank-2 Killing tensor relevant to the separability.
\par
The null geodesic equation of the spacetime is given by the Hamiltonian,
\begin{equation}
\mathcal{H}=\frac{1}{2}g^{ab}k_a k_b
=\frac{1}{2}\left[-f^{-1}(r)k_t^2+h^{-1}(r)k_r^2+r^{-2}\gamma^{ij}(x)k_ik_j\right]
=0,
\end{equation}
for a null geodesic tangent $k$, where $\gamma^{ij}$ is the inverse matrix of $\gamma_{ij}$.
Scaling the null vector $k$ as $k/E\to k$ by the constant of motion $E:=-k_t$, the equation can be rewritten as
\begin{equation}
-r^2f^{-1}(r)+r^2h^{-1}(r)k_r^2+\gamma^{ij}(x)k_ik_j=0.
\end{equation}
Since the first two terms are functions of $(r,k_r)$ and the last term is that of $(x^i,k_i)$, the equation is separated so that
\begin{equation}
\label{eq:null-geodesic-eq}
\frac{1}{2}\dot{r}^2+\widetilde{V}(B^2;r)=0,\;\;\; \widetilde{V}(B^2;r):=\frac{1}{2fh}\left(B^2fr^{-2}-1\right),
\end{equation}
where $\dot{r}=k^r$ is the derivative of $r$ with respect to an affine parameter $\lambda$ and $B^2$ is the separation constant having the relation
\begin{equation}
\label{eq:separation-const}
B^2:=\gamma^{ij}(x)k_ik_j\ge0.
\end{equation}
The variable $r$ of a null geodesic is separated and obeys this one-dimensional equation of motion.
\par
Now we consider a null geodesic $\gamma(\lambda)$ which satisfies $\dot{r}(\lambda)=0$ and $\ddot{r}(\lambda)=0$ for all $\lambda$.
From Eq.~\eqref{eq:null-geodesic-eq}, for $\gamma(\lambda)$, it is necessary and sufficient to satisfy $\widetilde{V}(B^2;r)=0$ and $\widetilde{V}'(B^2;r)=0$.
It is equivalent to give two of initial conditions for $\gamma(\lambda)$ at $\lambda=0$ by
\begin{eqnarray}
r(0)&=&r_{\mathrm{p}},\\
B^2&=&B^2_{\mathrm{p}},
\end{eqnarray}
where $r_{\mathrm{p}}$ and $B^2_{\mathrm{p}}$ are constants given by $\widetilde{V}(B^2_{\mathrm{p}};r_{\mathrm{p}})=0$ and $\widetilde{V}'(B^2_{\mathrm{p}};r_{\mathrm{p}})=0$ and the remaining initial conditions are arbitrary.
One can easily prove that the two conditions are equivalently expressed as
\begin{eqnarray}
r(0)&=&r_{\mathrm{p}},\\
\dot{r}(0)&=&0
\end{eqnarray}
according to Eq.~\eqref{eq:null-geodesic-eq}.
Furthermore, from the view point of the hypersurface $S_{r_{\mathrm{p}}}$, these are equivalent to
\begin{eqnarray}
\label{eq:init-in-srp}
\gamma(0)&\in& S_{r_{\mathrm{p}}},\\
\label{eq:init-tangent-srp}
\dot{\gamma}(0)&\in& T_{\gamma(0)}S_{r_{\mathrm{p}}}.
\end{eqnarray}
Thus, at every point $p$ on $S_{r_\mathrm{p}}$, every null geodesic satisfying the initial conditions~\eqref{eq:init-in-srp} and~\eqref{eq:init-tangent-srp} satisfies $\dot{r}=0$ and $\ddot{r}=0$, i.e., it remains tangent to $S_{r_\mathrm{p}}$.
According to Definition~\ref{definition:photonsurface} and~\ref{definition:r-photonsurface}, the hypersurface $S_{r_\mathrm{p}}$ is an $r$-photon surface.
\par
In summary, for any constant $r_\mathrm{p}$ given by the conditions $\widetilde{V}(B^2_{\mathrm{p}};r_{\mathrm{p}})=0$ and $\widetilde{V}'(B^2_{\mathrm{p}};r_{\mathrm{p}})=0$ with some constant $B^2_{\mathrm{p}}$, the hypersurface $S_{r_\mathrm{p}}$ is an $r$-photon surface.
In fact, we can obtain Eq.~\eqref{eq:V'=0} for $r_\mathrm{p}$ from these conditions, and therefore, the above argument is consistent with Proposition~\ref{proposition:r-psf}, which is derived from the geometrical analysis based on Theorem~\ref{theorem:equivalent-condition}.
The analysis based on the geodesic equation determines a photon surface and the null geodesics on it corresponding to the impact parameter $B_\mathrm{p}^2$ while the geometrical analysis giving Proposition~\ref{proposition:r-psf} does only a photon surface.
\par
The above analysis tells us that an $r$-photon surface is given by the equation in terms of the one-dimensional pseudopotential $V(r)$, Eq.~\eqref{eq:V'=0}, due to the separation of $r$ in the geodesic equation.
The separation of $r$ in the spacetime is due to the existence of a rank-$2$ Killing tensor, a symmetric tensor $K_{ab}$ satisfying the equation,
\begin{equation}
\nabla_{(c}K_{ab)}=0,
\end{equation}
where the brackets symmetrize the indices~\cite{textbook:wald}. 
The spacetime with the metric ansatz~\eqref{eq:metric-arbitrary} admits a rank-$2$ Killing tensor,
\begin{equation}
K_{ab}=r^4\gamma_{ab},
\end{equation}
where $\gamma_{ab}=\gamma_{ij}(dx^i)_a(dx^j)_b$.
(See Appendix~\ref{sec:comp-kt} for the computation of the Killing tensor equation for this $K_{ab}$.)
The Killing tensor is relevant to the separation of $r$.
Actually, it satisfies
\begin{equation}
K_{ab}k^ak^b=r^4\gamma_{ij}k^ik^j=\gamma^{ij}k_ik_j=B^2
\end{equation}
for any null geodesic tangent $k$.
Therefore, here we conclude that the spacetime can have photon surfaces in several cases regardless of the spatial symmetry because it admits the Killing tensor.
From this point of view, the spherically symmetric case of the spacetime, which is known to admit the photon surface (i.e., photon sphere) in many cases, is the case where the Killing tensor $K_{ab}$ is reducible to the sum of the products of the Killing vectors relevant to the spherical symmetry.

\section{Conclusion}
\label{sec:conclusion}
In this paper, we have investigated $r$-photon surfaces in the spacetime given by the metric ansatz, Eq.~(\ref{eq:metric-arbitrary}).
The ansatz is a general form of a warped spacetime as shown in Appendix~\ref{app:warped-product}.
The pseudopotential also implies that stable and unstable $r$-photon surfaces appear alternately as in the cases of light rings in spherically symmetric and rotationally symmetric spacetimes shown in Refs.~\cite{cunha_2017,cunha_2020}.
\par
We have first found that the pseudopotential $V(r)$ gives the radius $r$ and stability of an $r$-photon surface (Proposition~\ref{proposition:pseudo-potential}).
The local maxima correspond to unstable $r$-photon surfaces while the local minima correspond to stable ones.
It is remarkable that the stabilities of null geodesics on an $r$-photon surface depend on neither their directions nor positions on the surface even if the spacetime is not spatially symmetric.
\par
The $r$-photon surfaces indeed exist in the case where the spacetime is the electrovacuum solution to the Einstein equation with the cosmological constant.
Since the spacetime is a solution as far as the $(D-2)$-subspace $(\Sigma,\gamma)$ is an Einstein manifold, it implies that static photon surfaces exist in less or non-symmetric electrovacuum spacetimes.
Although many static photon surfaces have been found in highly symmetric spacetimes so far, our results imply that the existence of spatial Killing vectors is not crucial.
\par
We have also discussed the relation between photon surfaces and a Killing tensor.
Photon surfaces may exist if the null geodesic equation is well separable, and the separability is guaranteed by Killing vectors or Killing tensors.
In the present case, the warped spacetime we have investigated has a Killing tensor but does not have spatial Killing vectors in general.
Thus, here we conclude that a static photon surface may exist because of the Killing tensor rather than the spatial Killing vectors, or in other words, it does not necessarily require a high degree of spatial symmetry.
The existence of photon surfaces and the separability of the null geodesic equation has also been pointed out in Ref.~\cite{gibbons_2016}.
The authors investigated various types of the C-metric and found photon surfaces.
Although the C-metric does not admits spatial Killing vectors as many as a spherically symmetric spacetime, there exists a conformal Killing tensor, and therefore, the null geodesic equation is separable.
From this fact, one can further expect that, more generally, a conformal Killing tensor is crucial for photon surfaces.
\par
We finally make a remark about a possible extension of the above conclusion.
In the Kerr spacetime, there are spherical photon orbits (SPOs), i.e., orbits of constant radius with varying polar and azimuthal angles, and the SPOs of the same radius $r$ form a hypersurface~\cite{teo_2003}.
The hypersurface of constant $r$, which we denote here $S_r^{\mathrm{Kerr}}$, is similar to a photon surface in the sense that for some null vector $k\in T_pS_r^{\mathrm{Kerr}}$ at every point $p\in S_r^{\mathrm{Kerr}}$, there is a null geodesic which remains tangent to $S_r^{\mathrm{Kerr}}$.
Only the difference from the definition of a photon surface (Definition~\ref{definition:photonsurface}) is the part, ``for some null vector $k\in T_pS_r^{\mathrm{Kerr}}$."
Similarly to the fact that a photon surface is a totally umbilic hypersurface, the surface $S_r^{\mathrm{Kerr}}$ is called a partially umbilic hypersurface from the geometrical point of view according to Ref.~\cite{kobialko_2020}.
Remarkably, although the Kerr spacetime is of cohomogeneity two, the SPOs exist due to the separability and the relevant Killing tensor~\cite{benenti_1979,pappas}.
Therefore, we can expect that for generalized notions of photon surfaces or photon spheres, the existence of the Killing tensor would be crucial.
Specifically, it is interesting to investigate the relation between a Killing tensor and the generalizations of a photon sphere defined from the different points of view in Refs.~\cite{kobialko_2020,yoshino_tts,yoshino_dtts,siino_2019,shiromizu_2017}.
\begin{acknowledgments}
The authors are grateful to T. Harada, M. Kimura, T. Kokubu, and T. Katagiri for their fruitful discussions.
%
This work was supported by JSPS KAKENHI Grant No. JP19J12007 (Y.K.) and Grant-in-Aid for Early-Career Scientists~(JSPS KAKENHI Grant No.~JP19K14715) (T.I.) from the Japan Society for the Promotion of Science and the Rikkyo University Special Fund for Research (K.N.).
\end{acknowledgments}

\appendix
\section{STATIC WARPED SPACETIME}
\label{app:warped-product}
Here, we see that a spacetime $(M,g)$ with the metric $g$ given by Eq.~(\ref{eq:metric-arbitrary}) can be obtained as a generic static warped product of some class.
Consider a warped product of manifolds of the form,
\begin{equation}
M=M_1\times \mathcal{F}M_2,
\end{equation}
where $M_1$ is a two-dimensional Lorentzian manifold, $M_2$ is a $(D-2)$-dimensional Riemannian manifold, and $\mathcal{F}\colon M_1\to \mathbb R_{>0}$ is the warping function.
Letting $\{y^A\}$ and $\{x^i\}$ be coordinates on $M_1$ and $M_2$, respectively, we have
\begin{equation}
g=g_{AB}(y)dy^Ady^B+\mathcal{F}(y)\gamma_{ij}(x)dx^idx^j.
\end{equation}
Choosing $t\in\{y^A\}$ as the static time and $R\in\{y^A\}$ as the coordinate orthogonal to $t$, we have
\begin{equation}
g=g_{tt}(R)dt^2+g_{RR}(R)dR^2+\mathcal{F}(R)\gamma_{ij}(x)dx^idx^j.
\end{equation}
If we additionally assume that $\mathcal{F}'(R)\neq0$, we can transform the coordinate $R$ to $r$ defined by 
$r^2=\mathcal{F}(R)$.
Then, using the fact that $dr=d(\mathcal{F}^{1/2})=(1/2)\mathcal{F}^{-1/2}\mathcal{F}'dR$, we obtain
\begin{equation}
g=g_{tt}(R(r))dt^2+g_{rr}(R(r))dr^2+r^2\gamma_{ij}(x)dx^idx^j,
\end{equation}
where $g_{rr}(R(r))=(4r^2/(\mathcal{F}')^2)g_{RR}$.
Defining $-f(r):=g_{tt}(R(r))$ and $h(r):=g_{rr}(R(r))$, we obtain Eq.~(\ref{eq:metric-arbitrary}).

\section{COMPUTATION OF THE KILLING TENSOR EQUATION}
\label{sec:comp-kt}
We see that the tensor 
\begin{equation}
K_{ab}=r^4\gamma_{ab},
\end{equation}
where $\gamma_{ab}=\gamma_{ij}(dx^i)_a(dx^j)_b$, satisfies the Killing tensor equation
\begin{equation}
\nabla_{(c}K_{ab)}=0.
\end{equation}
The lhs is calculated as
\begin{eqnarray}
\nabla_{(\rho}K_{\mu\nu)}
&=&\partial_{(\rho}K_{\mu\nu)}
-\Gamma^\sigma{}_{(\rho\mu}K_{\nu)\sigma}
-\Gamma^\sigma{}_{(\rho\nu}K_{\mu)\sigma}
\nonumber\\
&=&\partial_{(\rho}K_{\mu\nu)}
-2\Gamma^\sigma{}_{(\rho\mu}K_{\nu)\sigma}
\nonumber\\
&=&\partial_{(\rho}K_{\mu\nu)}
-2\Gamma^l{}_{(\rho\mu}K_{\nu)l}
\nonumber\\
&=&\frac{1}{3}\left[\partial_{\rho}K_{\mu\nu}+\partial_{\mu}K_{\nu\rho}+\partial_{\nu}K_{\rho\mu}\right]
-\frac{2}{3}\left[\Gamma^l{}_{\rho\mu}K_{\nu l}+\Gamma^l{}_{\mu\nu}K_{\rho l}+\Gamma^l{}_{\nu\rho}K_{\mu l}\right]
\nonumber\\
&=&\frac{4}{3}r^3\left[\delta^r_{\rho}\gamma_{\mu\nu}+\delta^r_{\mu}\gamma_{\nu\rho}+\delta^r_{\nu}\gamma_{\rho\mu}\right]
+\frac{1}{3}r^4\left[\partial_{\rho}\gamma_{\mu\nu}+\partial_{\mu}\gamma_{\nu\rho}+\partial_{\nu}\gamma_{\rho\mu}\right]\nonumber\\
&&-\frac{2}{3}r^4\left[\Gamma^l{}_{\rho\mu}\gamma_{\nu l}+\Gamma^l{}_{\mu\nu}\gamma_{\rho l}+\Gamma^l{}_{\nu\rho}\gamma_{\mu l}\right].
\end{eqnarray}
All the terms vanish if one of the indices $\rho,\mu,\nu$ is $t$.
Thanks to the symmetrization of the indices, we need the calculations only for $(\rho,\mu,\nu)=(r,r,r),\ (r,r,i),\ (r,i,j),\ (i,j,k)$.
The Christoffel symbols we need are 
\begin{equation}
\Gamma^l{}_{rr}=0,\ \ \ \Gamma^l{}_{ri}=r^{-1}\delta^l_i,\ \ \ \ \Gamma^l{}_{ij}={}^\gamma\Gamma^l{}_{ij},
\end{equation}
where ${}^\gamma\Gamma^l{}_{ij}$ is the Christoffel symbols associated with $\gamma_{ij}$.
Then, for $(\rho,\mu,\nu)=(r,r,r)$, 
\begin{eqnarray}
\nabla_{(r}K_{rr)}
&=&\frac{4}{3}r^3\left[\delta^r_{r}\gamma_{rr}+\delta^r_{r}\gamma_{rr}+\delta^r_{r}\gamma_{rr}\right]
+\frac{1}{3}r^4\left[\partial_{r}\gamma_{rr}+\partial_{r}\gamma_{rr}+\partial_{r}\gamma_{rr}\right]\nonumber\\
&&-\frac{2}{3}r^4\left[\Gamma^l{}_{rr}\gamma_{r l}+\Gamma^l{}_{rr}\gamma_{r l}+\Gamma^l{}_{rr}\gamma_{r l}\right]\nonumber\\
&=&0.
\end{eqnarray}
For $(\rho,\mu,\nu)=(r,r,i)$,
\begin{eqnarray}
\nabla_{(r}K_{ri)}
&=&\frac{4}{3}r^3\left[\delta^r_{r}\gamma_{ri}+\delta^r_{r}\gamma_{ir}+\delta^r_{i}\gamma_{rr}\right]
+\frac{1}{3}r^4\left[\partial_{r}\gamma_{ri}+\partial_{r}\gamma_{ir}+\partial_{i}\gamma_{rr}\right]\nonumber\\
&&-\frac{2}{3}r^4\left[\Gamma^l{}_{rr}\gamma_{i l}+\Gamma^l{}_{ri}\gamma_{r l}+\Gamma^l{}_{ir}\gamma_{r l}\right]\nonumber\\
&=&0.
\end{eqnarray}
For $(\rho,\mu,\nu)=(r,i,j)$,
\begin{eqnarray}
\nabla_{(r}K_{ij)}
&=&\frac{4}{3}r^3\left[\delta^r_{r}\gamma_{ij}+\delta^r_{i}\gamma_{jr}+\delta^r_{j}\gamma_{ri}\right]
+\frac{1}{3}r^4\left[\partial_{r}\gamma_{ij}+\partial_{i}\gamma_{jr}+\partial_{j}\gamma_{ri}\right]\nonumber\\
&&-\frac{2}{3}r^4\left[\Gamma^l{}_{ri}\gamma_{j l}+\Gamma^l{}_{ij}\gamma_{r l}+\Gamma^l{}_{jr}\gamma_{i l}\right]
\nonumber\\
&=&\frac{4}{3}r^3\gamma_{ij}-\frac{2}{3}r^4\left[r^{-1}\delta^l_i\gamma_{jl}+r^{-1}\delta^l_j\gamma_{il}\right]
\nonumber\\
&=&0.
\end{eqnarray}
For $(\rho,\mu,\nu)=(i,j,k)$,
\begin{eqnarray}
\nabla_{(i}K_{jk)}
&=&\frac{4}{3}r^3\left[\delta^r_{i}\gamma_{jk}+\delta^r_{j}\gamma_{ki}+\delta^r_{k}\gamma_{ij}\right]
+\frac{1}{3}r^4\left[\partial_{i}\gamma_{jk}+\partial_{j}\gamma_{ki}+\partial_{k}\gamma_{ij}\right]\nonumber\\
&&-\frac{2}{3}r^4\left[\Gamma^l{}_{ij}\gamma_{k l}+\Gamma^l{}_{jk}\gamma_{i l}+\Gamma^l{}_{ki}\gamma_{j l}\right]
\nonumber\\
&=&\frac{1}{3}r^4\left[\partial_{i}\gamma_{jk}+\partial_{j}\gamma_{ki}+\partial_{k}\gamma_{ij}\right]
-\frac{2}{3}r^4\left[{}^\gamma\Gamma^l{}_{ij}\gamma_{k l}+{}^\gamma\Gamma^l{}_{jk}\gamma_{i l}+{}^\gamma\Gamma^l{}_{ki}\gamma_{j l}\right]
\nonumber\\
&=&\frac{1}{3}r^4\left[\partial_{i}\gamma_{jk}+\partial_{j}\gamma_{ki}+\partial_{k}\gamma_{ij}\right.
\left.
-\left(\partial_{j}\gamma_{ki}+\partial_{i}\gamma_{kj}-\partial_{k}\gamma_{ij}\right)
-\left(\partial_{k}\gamma_{ij}+\partial_{j}\gamma_{ik}-\partial_{i}\gamma_{jk}\right)\right.\nonumber\\
&&\left.
-\left(\partial_{i}\gamma_{jk}+\partial_{k}\gamma_{ji}-\partial_{j}\gamma_{ki}\right)
\right]
\nonumber\\
&=&0.
\end{eqnarray}
As a result, $\nabla_{(\rho}K_{\mu\nu)}=0$ and the tensor $K_{ab}$ is a Killing tensor.

%
\bibliography{rpsf_draft}

\end{document}